\documentclass[preprint,amsmath,amssymb,nofootinbib]{revtex4}
\usepackage{changes}
\usepackage{amsfonts}
\usepackage{amsmath,graphicx,color,epsfig}
\usepackage{epstopdf}
\usepackage{CJK}

\newcommand{\be}{\begin{equation}}
\newcommand{\ee}{\end{equation}}
\newcommand{\bea}{\begin{eqnarray}}
\newcommand{\eea}{\end{eqnarray}}
\newcommand{\ba}{\begin{array}}
\newcommand{\ea}{\end{array}}
\newcommand{\bi}{\begin{itemize}}
\newcommand{\ei}{\end{itemize}}

\newcommand{\nslash}{\kern 0.2 em n\kern -0.50em /}
\newcommand{\kslash}{\kern 0.2 em k\kern -0.45em /}
\newcommand{\qslash}{\kern 0.2 em q\kern -0.45em /}
\newcommand{\pslash}{\kern 0.2 em p\kern -0.50em /}
\newcommand{\rslash}{\kern 0.2 em r\kern -0.50em /}
\newcommand{\sslash}{\kern 0.2 em s\kern -0.50em /}
\newcommand{\Sslash}{\kern 0.2 em S\kern -0.50em /}
\newcommand{\Pslash}{\kern 0.2 em P\kern -0.50em /}
\newcommand{\Dslash}{\kern 0.2 em D\kern -0.65em /\kern 0.15em}
\newcommand{\lf}{\left}
\newcommand{\rg}{\right}

\begin{document}
\title{Confronting pentaquark photoproduction with new LHCb observations}

\author{Xu Cao}
\affiliation{Institute of Modern Physics, Chinese Academy of Sciences, Lanzhou 730000, China}
\affiliation{School of Nuclear Science and Technology, University of Chinese Academy of Sciences, Beijing 100049, China}

\author{Jian-Ping Dai{\footnote{Corresponding author: daijianping@ihep.ac.cn}}}
\affiliation{Experimental Physics Division, Institute of High Energy Physics, Chinese Academy of Sciences, Beijing 100049, China}
\affiliation{INPAC, Shanghai Key Laboratory for Particle Physics and Cosmology, MOE Key Laboratory for Particle Physics,
Astrophysics and Cosmology, Shanghai Jiao-Tong University, Shanghai, 200240, China}

\begin{abstract}

  The newly measurement of production fractions of $P_c$ states by LHCb collaboration have put restriction on their branching
  ratios of $J/\psi p$ decay, thus constraining their photoproduction in $\gamma p\to J/\psi p$ reaction. We show the tension
  between LHCb results and the current experiments in search of $P_c$ photoproduction. We also find that the present information
  of branching ratios of $P_c\to J/\psi p$ has already confronted sharply with the models which study the nature of $P_c$.

\end{abstract}

\maketitle


Very recently, the LHCb collaboration reported their new results of pentaquarks in $\Lambda_b^0\to J/\psi p K^-$ decay~\cite{Aaij:2019vzc},
soon after the first presentation at the Rencontres de Moriond QCD Conference~\cite{LHCb:newPc}, with nine times data sample
more than that used in previous analysis~\cite{Aaij:2015tga}. A new narrow pentaquark state, $P_c$(4312), was observed
with a statistical significance of $7.3~\sigma$, and the previously reported $P_c$(4440) resonance was resolved into two
narrower states, $P_c$(4440) and $P_c$(4457), where the statistical significance of this two-peak interpretation is $5.4\sigma$.
The measured resonance parameters are summarized in Table~\ref{tab:BRnewPc}. These findings motivate immediately
theoretical effort to study the mass spectrum and decay properties of these states~\cite{Chen:2019asm,Ali:2019npk,Chen:2019bip,
Guo:2019fdo,He:2019ify,Huang:2019jlf,Liu:2019tjn,Shimizu:2019ptd,Xiao:2019aya,Xiao:2019mst,Guo:2019kdc,Yamaguchi:2019seo,Valderrama:2019chc}, which update formerly extensive exploration~\cite{Wu:2010vk,Wu:2010jy,Wang:2011rga,Yang:2011wz,Wu:2012md,Yuan:2012wz,Xiao:2013yca,Shen:2016tzq,Lin:2017mtz,Huang:2018wed} and advance our understanding
of exotic candidates~\cite{Chen:2016qju,Guo:2017jvc,Liu:2019zoy}.
\begin{table}[b]
  \begin{center}
  \begin{tabular}{c|c|c|c}
  \hline\hline
    state       &            Mass [MeV]             &          Width [MeV]            & $\mathcal{R}$ [\%]  \\
  \hline
  $P_c(4312)^+$ & $4311.9 \pm  0.7 ^{+6.8}_{-0.6}$  & $9.8 \pm  2.7 ^{+3.7}_{-4.5}$   & $0.30 \pm 0.07^{+0.34}_{-0.09}$ \\
  $P_c(4440)^+$ & $4440.3 \pm  1.3 ^{+4.1}_{-4.7}$  & $20.6 \pm  4.9 ^{+8.7}_{-10.1}$ & $1.11 \pm 0.33^{+0.22}_{-0.10}$ \\
  $P_c(4457)^+$ & $4457.3 \pm  0.6 ^{+4.1}_{-1.7}$  & $6.4 \pm  2.0 ^{+5.7}_{-1.9}$   & $0.53 \pm 0.16^{+0.15}_{-0.13}$ \\
  \hline\hline
  \end{tabular}
  \end{center}
  \caption{The $P_c$ states in $\Lambda_b^0\to J/\psi p K^-$ decay observed by LHCb~\cite{Aaij:2019vzc,LHCb:newPc}.
  \label{tab:BRnewPc}}
\end{table}

\begin{figure}
  \begin{center}
  {\includegraphics*[width=10.0cm]{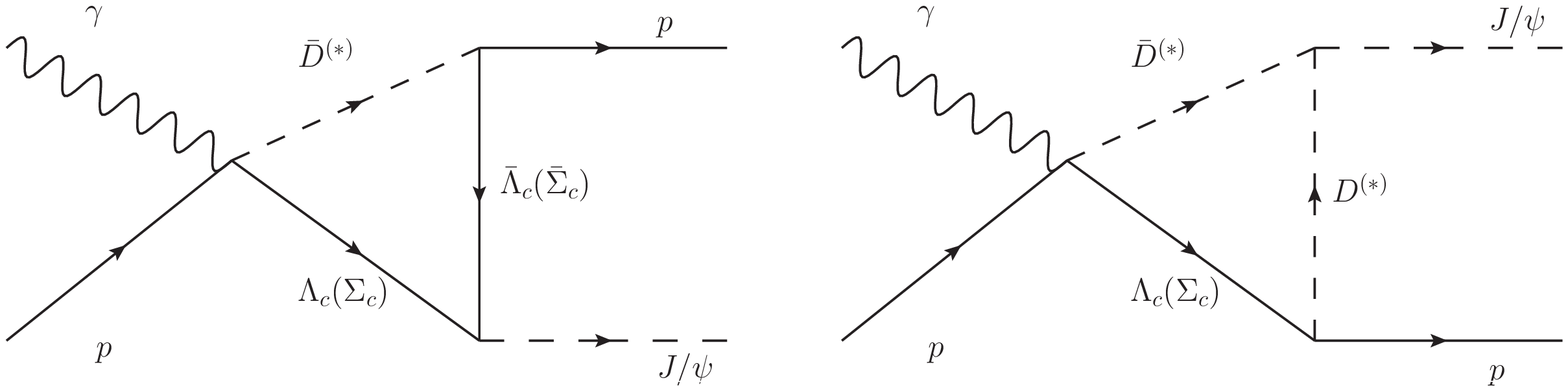}}
    \caption{Possible triangle diagrams for $\gamma p\to J/\psi p$. }
    \label{fig:JpsiTS}
  \end{center}
\end{figure}

The $P_c$ photoproduction in $\gamma p\to J/\psi p$ reaction has attracted wide interest soon after their discovery at
LHCb. Though triangle diagrams can be present in this reaction, e.g. in Figure~\ref{fig:JpsiTS}, it is hardly possible
to satisfy the on-shell conditions of the triangle singularity, which requires the masses of three particles in the triangle
to be on mass shell simultaneously. Thus these diagrams can not produce resonant-like structure and their contribution
is expected to be tiny. Hence the resonant structure produced in this reaction will be definitely genuine state. The calculation
in various models with moderate partial width of $P_c\to J/\psi p$ give sizable peaks upon the non-resonant $t$-channel
process in the cross section~\cite{Wang:2015jsa,Karliner:2015voa,Kubarovsky:2015aaa,Blin:2016dlf}. Several experimental
groups have put forward the corresponding proposal to study the pentaquark photoproduction, some of which are analyzing
the data and have released the preliminary results~\cite{GlueX:Pc,Ali:2019lzf,Meziani:2016lhg,Joosten:2018gyo}. The sensitivity
of these planned experiments is usually in the order of several percentage for the branching ratio $\mathcal{B}(P_c\to J/\psi p)$.
%

\begin{figure}
  \begin{center}
  {\includegraphics*[width=5.0cm]{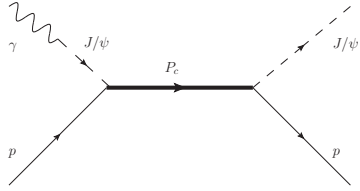}}
    \caption{The $P_c$ production in the $s$-channel of $\gamma p\to J/\psi p$. }
    \label{fig:Jpsipho}
  \end{center}
\end{figure}

The production of exotic state candidate $P_c$ in $s$-channel of $\gamma p\to J/\psi p$ can be generally written as~\cite{Karliner:2015voa,Kubarovsky:2015aaa}
\be \label{eq:sigmaR}
  \sigma_{P_c} = \frac{2\,J + 1}{2(2\,s_1 + 1)} \frac{4\,\pi}{k_{in}^2} \frac{\Gamma_{P_c}^2}{4} \frac{\mathcal{B}(P_c \to \gamma p) \, \mathcal{B}(P_c \to J/\psi p)}{(\sqrt{s} - M_{P_c})^2 + \Gamma_{P_c}^2/4},
\ee
where $s_1$ is the spin of initial proton. The $J$ is the total spin of $P_c$, $\sqrt{s}$ the center of mass (c.m.) energy,
and $k_{in}$ the magnitude of three momentum of initial states in c.m. frame. Figure~\ref{fig:Jpsipho} shows the corresponding
Feynman diagram. If assuming $P_c\to \gamma p$ is dominated by the vector meson (VMD), e.g. $J/\psi$ here, its branching
ratio $\mathcal{B}(P_c\to \gamma p)$ is proportional to $\mathcal{B}(P_c\to J/\psi p)$~\cite{Kubarovsky:2015aaa}:
\be \label{eq:txsBW}
\mathcal{B}(P_c\to \gamma p) = \frac{3\,\Gamma(J/\psi\to e^+e^-)}{\alpha M_{J/\psi}} \sum_L f_L \lf( \frac{k_{in}}{k_{out}} \rg)^{2L + 1}
\mathcal{B}_L (P_c\to J/\psi p), \,
\ee
where $\alpha$ is the fine structure constant, $L$ the quantum number of orbital excitation between $J/\psi$ and proton,
$k_{out}$ the magnitude of three momentum of final states in c.m. frame. The $f_L$, whose value can be found in Ref.~\cite{Kubarovsky:2015aaa},
is the fraction of decays $P_c\to J/\psi p$ in a relative partial wave $L$ that goes into transversally polarized $J/\psi$.
Thus the main uncertainty of the calculated cross section of $\gamma p\to J/\psi p$ is from the $\mathcal{B}(P_c\to J/\psi p)$.

In Table~\ref{tab:BRnewPc}, LHCb gives the updated value of branching fractions:
\be
\mathcal{R} = \frac{\mathcal{B}(\Lambda_b^0 \to P_c^+ K^-)\mathcal{B}(P_c^+ \to J/\psi p)}{\mathcal{B}(\Lambda_b^0 \to J/\psi p K^-)},
\ee
independent of the spin and parity of corresponding $P_c$. LHCb group has measured the $\mathcal{B}(\Lambda_b^0\to J/\psi p K^-)$
several years ago~\cite{Aaij:2015fea}, so one would obtain the value of $\mathcal{B}(P_c^+\to J/\psi p)$ if the branching
ratio of $\Lambda_b^0\to P_c^+ K^-$ is known. However, the $\mathcal{B}(\Lambda_b^0\to P_c^+ K^-)$ is dependent on the
nature of $P_c$ and even the intrinsic component of $\Lambda_b^0$~\cite{Hsiao:2015nna}, which is not fully determined yet,
therefore its value is hardly ever calculated and reported in the literature. As a compromise, we can only speculate a
range of $\mathcal{B}(\Lambda_b^0\to P_c^+ K^-)$ and $\mathcal{B}(P_c^+\to J/\psi p)$.


LHCb has given the results~\cite{Aaij:2015fea}:
\be \label{eq:LbJpLHCb}
\mathcal{B}(\Lambda_b^0\to J/\psi p K^-) = (3.17 \pm 0.04 \pm 0.07 \pm 0.34^{+0.45}_{-0.28}) \times 10^{-4},
\ee
whose accuracy is anticipated to be improved with the new LHCb data set. We will use the Particle Data Group (PDG) average
value~\cite{Tanabashi:2018oca}:
\be \label{eq:LbJpPDG}
\mathcal{B}(\Lambda_b^0\to J/\psi p K^-) = (3.2^{+0.6}_{-0.5}) \times 10^{-4}.
\ee
After combining it with $\mathcal{R}$ in Table~\ref{tab:BRnewPc}, we obtain
\be
\label{eqn:BRPcNEW}
\mathcal{B}(\Lambda_b\to P_c^+ K^-) \mathcal{B}(P_c^+\to J/\psi p)
= \begin{cases}
&(0.96^{+1.13}_{-0.39})\times10^{-6} \quad {\rm for}\quad P_c(4312)^+,\\
&(3.55^{+1.43}_{-1.24})\times10^{-6} \quad {\rm for}\quad P_c(4440)^+,\\
&(1.70^{+0.77}_{-0.71})\times10^{-6} \quad {\rm for}\quad P_c(4457)^+.
\end{cases}
\ee
GlueX at JLab Hall-D proposed an experiment to study the $P_c$ photoproduction and gave their result as~\cite{GlueX:Pc,Ali:2019lzf}:
\be \label{eq:GlueXPc}
\mathcal{B}(P_c^+\to J/\psi p) < 2.0\%.
\ee
The recently released specific values depend on the assumed spin-parity of $P_c$~\cite{Ali:2019lzf}, but they are all in
the above level and will not affect our conclusion. This upper limit is challenging the big partial widths of $J/\psi p$
in dynamically generated coupled-channel unitary approach~\cite{Wu:2010jy,Wu:2010vk} and the updated values with further
considering heavy quark symmetry~\cite{Xiao:2013yca,Xiao:2019aya}. This upper limit is also much smaller than the calculation
in molecular picture~\cite{Lin:2017mtz,He:2019ify,Xiao:2019mst,Chen:2019asm,Huang:2019jlf}. A very recent calculation in
the molecular scenario conclude that the $\mathcal{B}(P_c^+\to J/\psi p)$ for all $P_c$ is well above 10.0\%~\cite{Xiao:2019mst}.
In Table~\ref{tab:mwb}, the model calculations of $P_c$ partial width are summarized. As can be seen from the table, all
the calculated partial widths of $J/\psi p$ channel are above 1~MeV and the corresponding branching ratios are well above
several percent.\\
\begin{table}[ht]
  \setlength{\tabcolsep}{0.15cm}
  \begin{center}\caption{The mass (M), total width ($\Gamma$), and partial decay width ($\Gamma_{J/\psi p}$) of $J/\psi p$
  channel in various models are summarized for $P_c$. In Ref.~\cite{Xiao:2019mst}, the masses and total widths from new LHCb results~\cite{Aaij:2019vzc} are cited. In Ref.~\cite{Lin:2017mtz} the masses from old LHCb analysis~\cite{Aaij:2015tga} are used.}
  \begin{tabular}{cccc}\hline\hline
        Ref.                   & M(MeV) ($J^P$)                              & $\Gamma$(MeV)  & $\Gamma_{J/\psi p}$(MeV)  \\ \hline
   \cite{Wu:2010vk,Wu:2010jy}  & $4412$ ($\frac{1}{2}^-$, $\frac{3}{2}^-$)   & $47.3$    & $19.2$               \\

   \cite{Xiao:2013yca}         & $4262$ ($\frac{1}{2}^-$)                    & $35.6$    & $10.3$               \\
                               & $4410$ ($\frac{1}{2}^-$)                    & $58.9$    & $52.5$               \\
                               & $4481$ ($\frac{1}{2}^-$)                    & $57.8$    & $14.3$               \\
                               & $4334$ ($\frac{3}{2}^-$)                    & $38.8$    & $38.0$               \\
                               & $4417$ ($\frac{3}{2}^-$)                    & $8.2$     & $4.6$                \\
                               & $4481$ ($\frac{3}{2}^-$)                    & $34.7$    & $32.8$               \\

   \cite{Lin:2017mtz}          & $4380$ ($\frac{3}{2}^-$)                    & $144.3$   & $3.8$                \\
                               & $4450$ ($\frac{3}{2}^-$)                    & $139.8$   & $16.3$               \\
                               & $4450$ ($\frac{5}{2}^-$)                    & $46.4$    & $4.0$                \\

   \cite{Huang:2018wed}        & $4308$ ($\frac{1}{2}^-$)                    & $7.1$     & $1.2$                \\
                               & $4460$ ($\frac{1}{2}^-$)                    & $6.2$     & $3.9$                \\
                               & $4375$ ($\frac{3}{2}^-$)                    & $2.4$     & $1.5$                \\
                               & $4453$ ($\frac{3}{2}^-$)                    & $1.8$     & $1.5$                \\

   \cite{Xiao:2019mst}         & $4312$ ($\frac{1}{2}^-$)                    & $9.8$     & $3-7.5$              \\
                               & $4440$ ($\frac{1}{2}^-$, $\frac{3}{2}^-$)   & $20.6$    & $5.5-16$             \\
                               & $4457$ ($\frac{1}{2}^-$, $\frac{3}{2}^-$)   & $6.4$     & $2-4.5$              \\
 \hline\hline
 \end{tabular}  \label{tab:mwb}
 \end{center}
 \end{table}

With the upper limit in Eq.~(\ref{eq:GlueXPc}) we have,
\be \label{eqn:BRLbPcKNEW}
\mathcal{B}(\Lambda_b \to P_c^+ K^-)
>\begin{cases}
&(0.48^{+0.56}_{-0.20})\times10^{-4} \quad {\rm for}\quad P_c(4312)^+,\\
&(1.78^{+0.71}_{-0.62})\times10^{-4} \quad {\rm for}\quad P_c(4440)^+,\\
&(0.85^{+0.38}_{-0.36})\times10^{-4} \quad {\rm for}\quad P_c(4457)^+,
\end{cases}
\ee
which means that $P_c^+ K^-$ is a very important decay channel for $\Lambda_b^0$, whose branching fraction is at least in
the same level with $\mathcal{B}(\Lambda_b^0\to J/\psi p K^-)$ in Eq.~(\ref{eq:LbJpPDG}).

The values in Eq.~(\ref{eqn:BRPcNEW}) are nearly one order smaller than previous LHCb results~\cite{Aaij:2015fea}:
\be \label{eqn:BRPcOLD}
\mathcal{B}(\Lambda_b \to P_c^+ K^-) \mathcal{B}(P_c^+ \to J/\psi p)
=\begin{cases}
&(2.66\pm0.22\pm1.33^{+0.48}_{-0.38})\times10^{-5} \quad {\rm for}\quad P_c(4380)^+,\\
&(1.30\pm0.16\pm0.35^{+0.23}_{-0.18})\times10^{-5} \quad {\rm for}\quad P_c(4450)^+.
\end{cases}
\ee
If we sum $\mathcal{B}(\Lambda_b\to P_c^+ K^-) \mathcal{B}(P_c^+\to J/\psi p)$ of the $P_c(4440)^+$ and $P_c(4457)^+$ in
Eq.~(\ref{eqn:BRPcNEW}), we find that it is roughly compatible with that of $P_c(4450)^+$ in Eq.~(\ref{eqn:BRPcOLD}) within
two standard deviations. This is consistent with the observation that $P_c(4450)^+$ is decomposed into $P_c(4440)^+$ and
$P_c(4457)^+$ states in the new LHCb data sample. If regardless of this problem at the moment, we adopt these values in
Eq.~(\ref{eqn:BRPcOLD}) and GlueX's upper limit in Eq.~(\ref{eq:GlueXPc}), it means that,
\be
\label{eqn:BRLbPcKOLD}
\mathcal{B}(\Lambda_b \to P_c^+ K^-)
>\begin{cases}
&(1.33^{+0.72}_{-0.70})\times10^{-3} \quad {\rm for}\quad P_c(4380)^+,\\
&(0.65^{+0.22}_{-0.21})\times10^{-3} \quad {\rm for}\quad P_c(4450)^+,
\end{cases}
\ee
confronting with values in Eq.~(\ref{eqn:BRLbPcKNEW}). In this case it is expected that $\mathcal{B}(\Lambda_b^0\to P_c^+ K^-)$
is at least in the same level of $\mathcal{B}(\Lambda_b^0\to \Lambda_c^+\pi^-)$ and $\mathcal{B}(\Lambda_b^0\to \Lambda_c^+\pi^+\pi^-\pi^-)$,
only smaller than the leptonic branching decay ratio of $\Lambda_b^0$, which is in the value of several percentage~\cite{Tanabashi:2018oca}.
{This deduction challenges our understanding of $\Lambda_b^0$ properties. Furthermore, due to the narrow width of $P_c$,
if $\mathcal{B}(\Lambda_b\to P_c^+ K^-)$ is big as indicated in Eq.~(\ref{eqn:BRLbPcKOLD}), it would be relatively easy
to observe it in other decay modes, whose branching ratios are much larger than that of $J/\psi p$ decay, such as $\bar D\Sigma_c$,
$D^0\Lambda_c$.} Therefore the old LHCb results in Eq.~(\ref{eqn:BRPcOLD}) are confronting sharply with GlueX's
upper limit in Eq.~(\ref{eq:GlueXPc}) and unreasonable. A reasonable and very loose upper limit of $\mathcal{B}(\Lambda_b\to P_c^+ K^-)$
would be $10^{-3}$, which is consistent with Eq.~(\ref{eqn:BRLbPcKNEW}). Thus from Eq.~(\ref{eqn:BRPcNEW}) we have:
\be
\label{eqn:BRPcJplowlimit}
2\%>\mathcal{B}(P_c^+ \to J/\psi p)
>\begin{cases}
&(0.96^{+1.13}_{-0.39})\times10^{-3} \quad {\rm for}\quad P_c(4312)^+,\\
&(3.55^{+1.43}_{-1.24})\times10^{-3} \quad {\rm for}\quad P_c(4440)^+,\\
&(1.70^{+0.77}_{-0.71})\times10^{-3} \quad {\rm for}\quad P_c(4457)^+.
\end{cases}
\ee
As a result, we roughly expect $\mathcal{B}(P_c^+\to J/\psi p)$ for all $P_c$,
\be
\label{eqn:BRPcJplimit}
2\% > \mathcal{B}(P_c^+\to J/\psi p) > 0.05\%,
\ee
which is a tight constrain at the moment. A naive postulation is that $\mathcal{B}(\Lambda_b\to P_c^+ K^-)$ shall be not
bigger than $\mathcal{B}(\Lambda_b^0\to J/\psi p K^-)$ in Eq.~(\ref{eq:LbJpLHCb}). If we match this postulation and use
an upper limit $10^{-4}$ for $\mathcal{B}(\Lambda_b\to P_c^+ K^-)$, the lower limit of $\mathcal{B}(P_c^+\to J/\psi p)$
would be 0.5\%. However, we will adopt the conservative range in Eq.~(\ref{eqn:BRPcJplimit}).

The obtained lower bound of $\mathcal{B}(P_c^+\to J/\psi p)$ in Eq.~(\ref{eqn:BRPcJplimit}) is independent of the spin
and parity, because they are deduced from the model independent value from LHCb~\cite{Aaij:2019vzc}. The upper bound in
Eq.~~(\ref{eqn:BRPcJplimit}), inferred from recent GlueX data of $\gamma p\to J/\psi p$, depends on the spin $J$ of $P_c$.
But as can be seen in Eq.~\ref{eq:sigmaR}, the total production cross section of $\gamma p\to P_c \to J/\psi p$ reaction
is proportional to $2J + 1$. As a result, this upper bound would remain to be several percent if $P_c$ has different $J$,
hence it is not very sensitive to spin $J$ of $P_c$. Particularly, if the spin of $P_c(4440)$ ($P_c$(4457)) state is assigned
to be 3/2 (1/2), the cross section will increase (decrease) by two times than that of 1/2 (3/2) assignment. At present
stage, uncertainties in this level is tolerable.

This upper bound is also dependent on the vector meson model (VMD), which is used in the postulation of Eq.~(\ref{eq:txsBW}).
One of debates of this model is that the vector meson in the $\gamma-V$ vertex is off shell, so a squared form factor
$\mathcal{F}_V^2(q^2=0)$ shall be further introduced in Eq.~(\ref{eq:txsBW}), e.g. in the form of
\be \label{eq:ffvmd}
\mathcal{F}_V(q^2) = \frac{\Lambda_V^4}{\Lambda_V^4+(q^2-m_V^2)^2}.
\ee
Other form factors can be used. Anyway, $\mathcal{F}^2(q^2=0)$ is an undermined constant, as can be seen above. This is
closely related to another debate: which kind of vector meson - light ($\rho$, $\omega$, $\phi$) or heavy ($J/\psi$) -
dominates $\gamma-V$ vertex~\cite{Wu:2019adv}, critically depending on the magnitudes of cut-off $\Lambda_V$. The light
meson is predominant with $\Lambda_V \sim m_\rho$. Since the partial width of ${J/\psi p}$ is much bigger than other $V p$
as expected by nearly all models, if we insist that the value of $\Lambda_V$ shall be in the scale of the produced vector
meson $m_{J/\psi}$ in final states, the $J/\psi$ meson is dominant in $\gamma-V$ vertex, in line with our above consideration.
Then the upper limit 2\% in Eq.~(12) shall be changed to:
\be
\frac{2\%}{\mathcal{F}_{J/\psi}^2(q^2=0)},
\ee
which is sensitive to the choice of $\Lambda_V$. For example, it is 8.0\% with $\Lambda_V = m_{J/\psi}$ and 23.7\% with
$\Lambda_V = 0.8 m_{J/\psi}$. As a result, the large partial width of ${J/\psi p}$ in many models could be incorporated
into the absence of $P_c$ in $\gamma p\to J/\psi p$ data. Unfortunately, this form factor, originated from non-perturbative
effect, is hard to calculate and also can not be determined from other process. Therefore whether this form factor should
be added is under discussion.


Several models with molecular picture prefer the assignment of $1/2^-$, $1/2^-$, and $3/2^-$ for $P_c(4312)^+$, $P_c(4440)^+$,
and $P_c(4457)^+$, respectively~\cite{Wu:2012md,He:2019ify,Xiao:2019aya,Xiao:2019mst,Chen:2019asm,Huang:2019jlf}. This
assignment is supported by QCD sum rule~\cite{Chen:2019bip} but different from the interpretation of hidden-charm
diquark-diquark-antiquark baryons~\cite{Ali:2019npk}. If the first two $P_c$ states have the same spin-parity as expected
by most of the models, one may firstly assume that their $\mathcal{B}(\Lambda_b\to P_c^+ K^-)$ are roughly the same. From
Table~\ref{tab:BRnewPc}, we have
\be \label{eq:PcBr}
\mathcal{B}(P_c(4440)^+ \to J/\psi p): \mathcal{B}(P_c(4312)^+ \to J/\psi p) =  1: 0.27^{+0.32}_{-0.14}.
\ee
If at the moment we neglect the small difference of phase space from $P_c$ mass and assume the decay of two $P_c$ states
proceeding in the same partial wave in Eq.~(\ref{eq:txsBW}) (usually the lowest partial wave is dominant), the $P_c$ photoproduction
cross section $\sigma_{P_c}$ at the peak position is,
\be \label{sigmaPc1Pc2}
\sigma_{P_c(4440)^+} : \sigma_{P_c(4312)^+} \simeq 1: 0.07^{+0.18}_{-0.08}.
\ee
Apparently above values imply that $P_c(4440)$ is much easier to be found in $\gamma p\to J/\psi p$. {So if we do not find
$P_c(4440)^+$ in the $\gamma p\to J/\psi p$} and use the GlueX's upper limit in Eq.~(\ref{eq:GlueXPc}), the upper limit
of $\mathcal{B}(P_c(4312)^+\to J/\psi p)$ would be 0.54$^{+0.65}_{-0.28}$\%. Therefore by taking the uncertainties into
account we expect roughly that:
\bea \label{eqn:BRPcJp4312}
1.2\%>\mathcal{B}(P_c(4312)^+ \to J/\psi p)>0.05\%, \\
\label{eqn:BRPcJp4440}
2\%>\mathcal{B}(P_c(4440)^+ \to J/\psi p)>0.2\%.
\eea
As stated above, above conclusion, based on the fact that we do not find $P_c(4440)^+$ in the $\gamma p\to J/\psi p$ with present precision, is dependent on two assumptions:
\begin{itemize}
\item[(1)] The production mechanism of $P_c(4312)$ and $P_c(4440)$ is the same in the $\Lambda_b$ decay;

\item[(2)] The decay of two $P_c$ states to $J/\psi p$ is proceeding in the same partial wave,
\end{itemize}
both of which depend critically on the nature of $P_c(4312)$ and $P_c(4440)$. The first one about the $P_c$ production
is rarely studied by the models, though their decay is explored widely. But Eq.~(\ref{eq:PcBr}) would not change much if
all $P_c$ are of same nature, e.g. molecular states, considering $\Lambda_b\to P_c^+ K^-$ is weak decay and shall be of
the same magnitude for all $P_c$. The second one is trivially understandable because of usually the dominance of the lowest
partial wave. Several recent papers have an interesting finding that spin parity assignments for $P_c$(4440) and $P_c$(4457)
is sensitive to the one-pion exchange potential~\cite{Liu:2019tjn,Yamaguchi:2019seo,Valderrama:2019chc}. Consequently,
Eq.~(\ref{sigmaPc1Pc2}) would have the difference with a factor of $2J + 1$ if $P_c$'s have different spins, as discussed
below Eq.~(\ref{eqn:BRPcJplimit}). So the upper limit in Eq.~(\ref{eqn:BRPcJp4312}) and the lower limit in Eq.~(\ref{eqn:BRPcJp4440})
would have a difference with a factor of $2J + 1$, which is not big considering the present precision of our knowledge
of these branching ratios. As a result, Eqs.~(\ref{eq:PcBr}-\ref{eqn:BRPcJp4440}) is more sensitive to the internal structure
of these $P_c$. For example, if one of them is of molecular state while another one is tetraquark, then Eqs.~(\ref{eq:PcBr}-\ref{eqn:BRPcJp4440})
would change drastically. We have to admit this possibility is not excluded yet at present. In other words, if Eqs.~(\ref{eq:PcBr}-\ref{eqn:BRPcJp4440})
and similar relations for $P_c$ with other spin-parity is questioned by the data of $\gamma p \to J/\psi p$ in future,
then the nature of these $P_c$ would be quite different.

Eq.~(\ref{eqn:BRPcJp4312}) tells us that $P_c(4312)$ photoproduction is out of reach of current cross section measurements,
which is only sensitive to a few percentage of $\mathcal{B}(P_c\to J/\psi p)$. More specifically, if the role of $P_c(4312)$
is observed in $\gamma p\to J/\psi p$ by experiment, then the first assumption above would be most doubtful and the nature
of $P_c(4312)$ and $P_c(4440)$ are quite different. For $P_c(4440)$ and $P_c(4457)$, the situation is more complicated
due to the closeness of two states. If the experiments do not find any peak around their mass, one possibility is that
their couplings to $J/\psi p$ is really inconspicuous as shown by the lower limit in Eq.~(\ref{eqn:BRPcJp4440}) and Eq.~(\ref{eqn:BRPcJplimit}).
However, another possible reason may be the destructive interference between $P_c(4440)$ and $P_c(4457)$, even they both
have strong coupling to $J/\psi p$.

\begin{figure}
  \begin{center}
  {\includegraphics*[width=10.0cm]{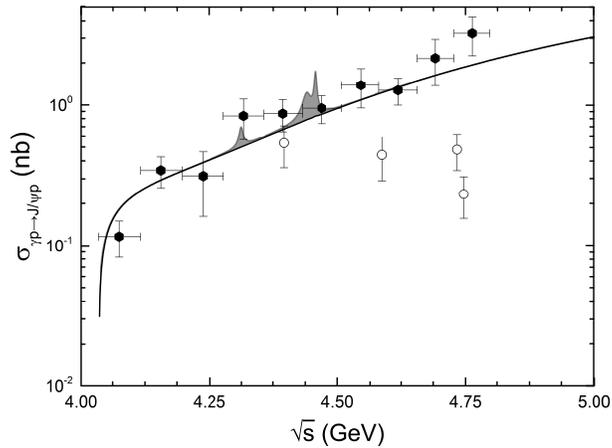}}
    \caption{The cross section of $\gamma p\to J/\psi p$ as a function of $\sqrt{s}$. The solid circles represent the GlueX
    data~\cite{Ali:2019lzf}, and the open circles is the old data from the compilation in Ref.~\cite{Martynov:2002ez}.
    The solid curve is from the non-resonant contribution parameterized with the soft dipole pomeron model~\cite{Martynov:2002ez},
    and the grey band represents the contribution of $P_c$ state. }
    \label{fig:Jpsiphoto}
  \end{center}
\end{figure}
The cross section of $\gamma p\to J/\psi p$ as a function of $\sqrt{s}$ is depicted in Figure~\ref{fig:Jpsiphoto}. The
non-resonant contribution, which is represented with the solid curve, is parameterized with the soft dipole pomeron model~\cite{Martynov:2002ez}.
The resonant contribution marked by the grey band is calculated by Eq.~(\ref{eq:sigmaR}) in consideration of the range
of $\mathcal{B}(P_c\to J/\psi p)$ in Eq.~(\ref{eqn:BRPcJp4312}) for $P_c(4312)$, Eq.~(\ref{eqn:BRPcJp4440}) for $P_c(4440)$
and Eq.~(\ref{eqn:BRPcJplimit}) for $P_c(4457)$, respectively. Here we use the central values of the parameters of $P_c$
states in Table~\ref{tab:BRnewPc} and do not consider the interference among various contributions, which is premature
to consider in the calculation. On one hand, the branching fraction of $P_c\to J/\psi p$ have bigger uncertainties as explored
in this paper. Also because of the unknown relative phases between any two different amplitudes, it is difficult to give
a reliable quantitative description of the interference. On the other hand, the role of interference is more explicit in
differential cross-sections and polarization observables, rather than the total cross sections. Take the $t$-dependent
cross sections for example, the resonant contribution is important for all $t$ range, while $t$-channel Pomeron exchange
and $u$-channel contribute to forward angles (small $|t|$) and backward angles (bit $|t|$), respectively. GlueX's data
follow with $t$-channel Pomeron behavior, with limited precision in big $|t|$ range. As can be seen, it is really challenging
to hunt for such narrow $P_c$ states in their photoproduction.

The model calculation showed that $P_c$ could be clearly evident in differential cross sections with $\mathcal{B}(P_c\to J/\psi p) = 5\%$~\cite{Wang:2015jsa}. The Hall-C at JLab is analyzing the new data of $\gamma p\to J/\psi p$ and expected to release the results soon. It can identify the $P_c$ photoproduction if the partial branching ratios of $J/\psi p$ decay are of several percentage~\cite{Joosten:2018gyo,Meziani:2016lhg}.
If they confirm the upper limit of GlueX at Hall-D in Eq.~(\ref{eq:GlueXPc}), then it is not easy to study the $P_c$
peaks in the unpolarized cross section of $\gamma p\to J/\psi p$, concerning that the cross section is proportional to $\mathcal{B}^2(P_c\to J/\psi p)$ in view of Eq.~(\ref{eq:sigmaR}) and Eq.~(\ref{eq:txsBW}).
Whereas the measurement of polarization asymmetries would still be encouraged. With the lower limit in Eq.~(\ref{eqn:BRPcJplimit}), Eq.~(\ref{eqn:BRPcJp4312}) and
Eq.~(\ref{eqn:BRPcJp4440}), we optimistically expect that $P_c$, at least for $P_c(4440)$, is noticeable in the polarization
observables if they are real resonant states, though they would not be obvious in the cross sections. The quantum numbers of $P_c$ and possible complex interference would be also differentiated by polarization measurements.

In a short conclusion, based on the branching ratios and fractions measured by LHCb and GlueX collaborations, we give a confined
range of $\mathcal{B}(P_c\to J/\psi p)$. The small $\mathcal{B}(P_c\to J/\psi p)$ are confronting sharply with the up-to-date
data of cross section of $P_c$ photoproduction. It is anticipated that the polarization observables in $\gamma p\to J/\psi p$ would
be more appreciate for searching for $P_c$ photoproduction and determining the assignment of their spin parity, in light of
the sensitivity of JLab experiments and limited magnitude of $\mathcal{B}(P_c\to J/\psi p)$. Other final states, e.g. $\bar{D}^0\Lambda_c$ in photoproduction, would be also crucial for looking for the $P_c$ states. We would like to address that the couplings of $P_c \to J/\psi p$ will disentangle various models of the $P_c$, some of which have given the calculated values but well above the present range. Some models predicted different assignment of spin-parity of $P_c$ from those used to speculate Eq.~(\ref{eqn:BRPcJp4312}) and Eq.~(\ref{eqn:BRPcJp4440}). We can similarly infer the corresponding range for them to constrain the model parameters.
On the other hand, we have limited information on $P_c$ production mechanism, especially the $\Lambda_b\to P_c^+ K^-$. Our given range of $\mathcal{B}(\Lambda_b\to P_c^+ K^-)$ in Eq.~(\ref{eqn:BRLbPcKNEW}) is relatively big, which needs experimental confirmation and an appropriate interpretation. Hence it needs more theoretical and experimental attention on this aspect in the future.

\begin{acknowledgments}

One of author (X. C.) would like to thank Prof. E.~Martynov for useful communication about the experimental data and the
soft dipole pomeron model in Fig.~\ref{fig:Jpsiphoto}. Useful discussions with Dr. Zhi Yang, Jia-Jun Wu, Xiao-Hai Liu, and Ju-Jun Xie
are grateful acknowledged. We thank Prof. Cheng-Ping Shen for the useful comments on the manuscript. This work was supported
by the National Natural Science Foundation of China (Grants Nos. 11405222 and 11505111) and the Key Research Program of
Chinese Academy of Sciences (Grant NO. XDPB09).

\end{acknowledgments}

\end{document}